\documentclass[aps, prx, twocolumn, amssymb, amsmath, superscriptaddress]{revtex4-1}

\usepackage{bm}
\usepackage{times}
\usepackage{graphicx}
\usepackage[colorlinks,citecolor=blue,linkcolor=blue]{hyperref}
\usepackage{array}
\usepackage{float}
\usepackage{colortbl}
\usepackage{multirow}
\usepackage{tabularx}

\begin{document}

\title{The Doping effect of Chalcogen on the Two-Dimensional \\
 Ferromagnetic Material Chromium Tribromide}
\author{Yanfeng Ge}
\affiliation{Technology and Key Laboratory for Microstructural Material Physics of Hebei Province, School of Science, Yanshan University, Qinhuangdao, 066004, People's Republic of China}

\author{Qiaoqiao Li}
\affiliation{Technology and Key Laboratory for Microstructural Material Physics of Hebei Province, School of Science, Yanshan University, Qinhuangdao, 066004, People's Republic of China}

\author{Wenhui Wan}
\affiliation{Technology and Key Laboratory for Microstructural Material Physics of Hebei Province, School of Science, Yanshan University, Qinhuangdao, 066004, People's Republic of China}

\author{Jian-Min Zhang}
\affiliation{Fujian Provincial Key Laboratory of Quantum Manipulation and New Energy Materials, College of Physics and Energy, Fujian Normal University, Fuzhou, 350117, People's Republic of China}

\author{Wenhui Xie}
\affiliation{Department of Physics, East China Normal University, Shanghai 200062, People's Republic of China}

\author{Yong Liu}\email{yongliu@ysu.edu.cn}
\affiliation{Technology and Key Laboratory for Microstructural Material Physics of Hebei Province, School of Science, Yanshan University, Qinhuangdao, 066004, People's Republic of China}
\date{\today}

\begin{abstract}
Recently the discovery of magnetic order in two-dimensional monolayer chromium trihalides opens the new research field in two-dimensional materials. We use first-principles calculations to systematically examine the doping effect of chalcogen on CrBr$_3$. In the case of S-doping, four stable configurations, Cr$_2$Br$_5$S, Cr$_2$Br$_4$S$_2$-A, Cr$_2$Br$_4$S$_2$-B and Cr$_2$Br$_3$S$_3$-A, are predicted to be ferromagnetic semiconductors. It is found that the new bands appearing in the original bandgap are made up of S-p and Cr-d-e$_g$ orbits, lead to the obvious reduce of bandgap and the enhanced optical absorption in the visible range. Due to the decrease of valence electron after chalcogen doping, the magnetic moment also decreases with the increase of S atoms, and the character of ferromagnetic semiconductor is always hold in a wide range of strain. The results shown that monolayer CrBr$_3$ with chalcogen doping supply a effectual way to control the magnetism and extend the optoelectronic applications.

\end{abstract}

\maketitle

\section{Introduction}

In the last decade, two-dimensional (2D) materials~\cite{Novoselov2005,Zhang2005,Geim2013,Kim2015,Xi2016,Saito2016} have attracted a great deal of attention also been one of the most exciting research fields due to the wealth of physics and promising applications. Besides the earliest graphene, various 2D materials also include hexagonal boron nitride, silicene, transition metal dichalcogenides, MXene and so on. However, the absent of intrinsic magnetism in the most of the available 2D materials with pristine form limits their applications in spintronics and spin-based electronics, particularly information technology. As a consequence, 2D ferromagnetic materials with the combination of large spin polarization and high Curie temperature (T$_C$) are of particular importance and interest. The simple and direct means to achieve magnetic in 2D materials is dopants of magnetic atoms. However, doping not only destroys the perfect crystalline order, but also the low solubility and surface clustering~\cite{Wang2005,Dietl2010} are the serious problems.
Although there are some other strategies to magnetize~\cite{Han2014}, the long-range magnetic order is rarely observed experimentally in 2D materials. For the theoretical reason, Mermin and Wagner theorem demonstrate the strong fluctuations lead to the absence of long range magnetic order in spin-rotational invariant systems with short range exchange interactions.

Fortunately, the magnetocrystalline anisotropy make some materials be not limited by the Mermin-Wagner theorem. The recent reports of magnetic order in different 2D crystals (Cr$_2$Ge$_2$Te$_6$, CrI$_3$ and FePS$_3$)~\cite{Gong2017,Huang2017,Wang2016,Lee2016,Wang2018} mark a milestone in the research of 2D magnetic materials and explore the 2D magnetic physics, indispensable for the low-dimensional spintronics.
The small cleavage energy in the ferromagnetic van der Waals (vdW) bulk CrI$_3$ is propitious to the mechanically exfoliated monolayer structure, prevalent in other common 2D materials. The magnetization experiments show the bulk crystal has the out-of-plane ferromagnetic (FM) order below T$_C$ = 68 K~\cite{McGuire2015,Hansen1959,Dillon1965} with weak interlayer coupling. FM order is still residing in the monolayer structure with the lower T$_C$ of 45 K~\cite{Huang2017}. But the suppressed magnetization occurs in the bilayer CrI$_3$ due to the metamagnetic effect~\cite{Stryjewski1977}. The numerous publications with regard to CrI$_3$ emerge in order to study and manipulate the magnetic properties~\cite{Klein2018,Song2018,Frisk2018,PeihengJiang2018,ChengxiHuang2018}. The temperature-dependent magnetic anisotropy~\cite{Richter2018} of 2D CrI$_3$ is contributed from the ferromagnetic super-exchange across the anisotropic Cr-I-Cr bonds and the weak single ion anisotropy of Cr atom~\cite{Lado2017}. Furthermore, the electric field~\cite{Huang2018}, electrostatic doping~\cite{Jiang2018} and strain~\cite{Zheng2018} are found to control the 2D magnetism of CrI$_3$. Besides, the homogeneous materials CrBr$_3$ and CrCl$_3$ also induce the research interests~\cite{Tsubokawa1960,Wang2011,Liu2016,Zhang2015,Abramchuk2018,Li2018,McGuire2017} and the bulk CrBr$_3$ in particular has been studied as ferromagnetic semiconductor with T$_c$ = 31 K long ago~\cite{Tsubokawa1960}.
And the in-plane multiferroicity~\cite{ChengxiHuang2018}, such as ferromagnetic and ferroelectricity order, are also discovered in the charged monolayer CrBr$_3$. The mixed halide series CrCl$_{3-x}$Br$_x$ realizes the regulation of exchange anisotropy~\cite{Abramchuk2018} and can be applied to the ultrathin magneto-optical devices~\cite{Zhong2017,Wang2017,Seyler2018}.

\begin{figure*}
\centerline{\includegraphics[width=0.95\textwidth]{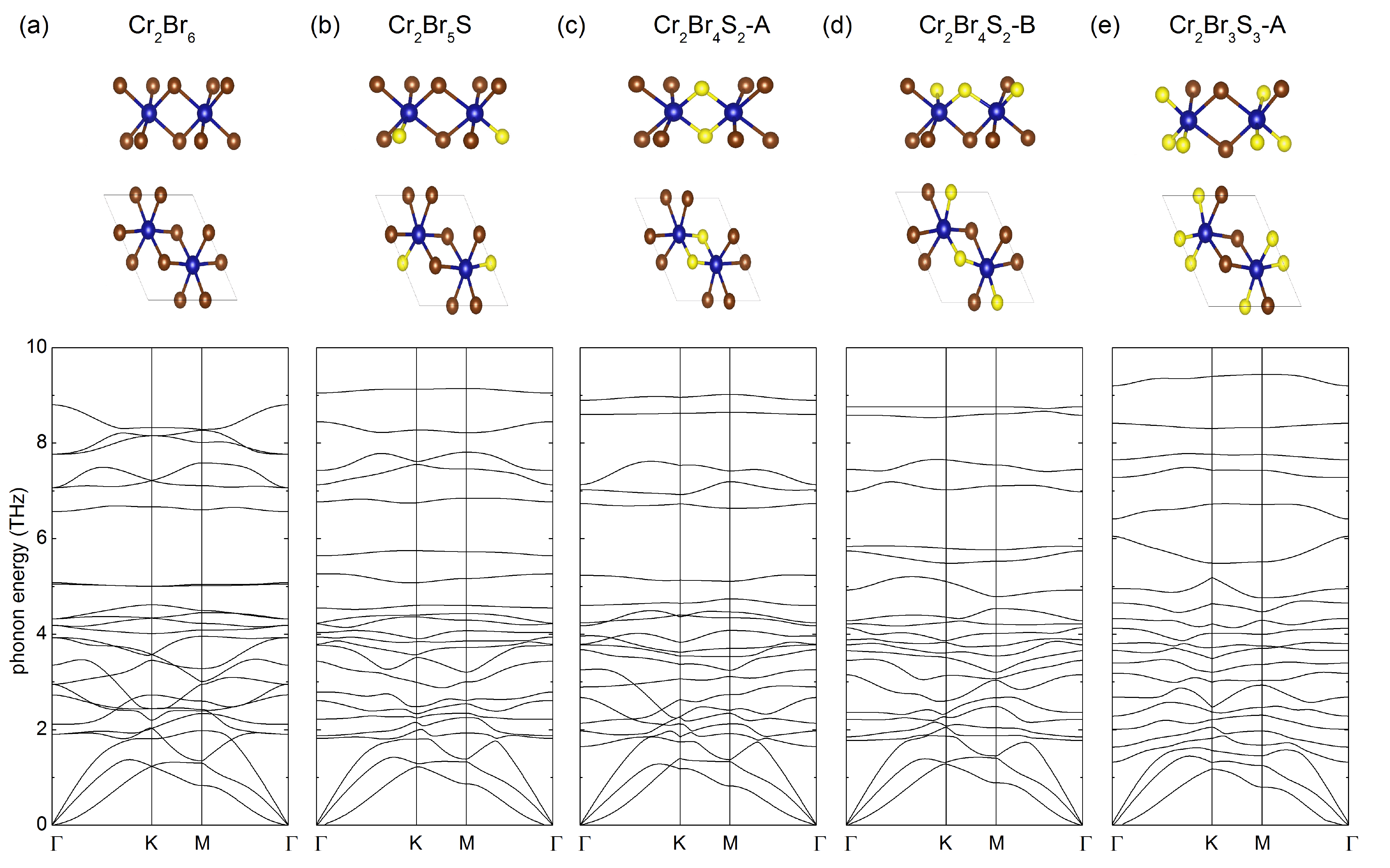}}
\caption{{\bf Crystal structures and phonon spectra of Cr$_2$Br$_{6-x}$S$_x$.} (a) Crystal structure (upper panel) and phonon spectrum (bottom panel) of pristine CrBr$_3$, whose primitive cell has two Cr atoms and six Br atoms (Cr$_2$Br$_6$). In each primitive cell, Br atoms substituted by (b) one S atom: Cr$_2$Br$_5$S, (c) two S atoms: Cr$_2$Br$_4$S$_2$-A, (d) two S atoms : Cr$_2$Br$_4$S$_2$-B and (e) three S atoms: Cr$_2$Br$_3$S$_3$-A.
\label{fig:phonon}}
\end{figure*}

In the present work, we focus on the 2D semiconductor CrBr$_3$ and study the doping effect of chalcogen (S, Se and Te), in consideration of ferromagnetic metal in chromium chalcogens~\cite{Lishuo2018,CongWang2018} with the same crystal structure as CrI$_3$.
Except the prototype structure of CrBr$_3$, Six types of doping structure have been constructed with doping concentration of 1 $\sim$ 3 chalcogenide atoms per primitive cell. In the case of Cr$_2$Br$_{6-x}$S$_x$, four of the structures have dynamic stability, verified by the phonon spectra. They are as follows: Cr$_2$Br$_5$S, Cr$_2$Br$_4$S$_2$-A, Cr$_2$Br$_4$S$_2$-B and Cr$_2$Br$_3$S$_3$-A. By calculating spin-polarized electronic structures, the four stable configurations are also confirmed as ferromagnetic semiconductor, coordination with CrBr$_3$. However, there new bands appears in the bandgap of Cr-d-t$_{2g}$ orbit and the number of new bands increases linearly with S atomic content, which impact the global bandgap significantly. Due to the reduce of valence electron, magnetic moment also decreases under the doping condition, such as 3 $\mu_{\rm B}$ per Cr$_2$Br$_{3}$S$_3$-A primitive cell. These results suggest that monolayer CrBr$_3$ with chalcogen doping is also robust intrinsic ferromagnetic semiconductor. Moreover, the new energy band can also absorb the low-energy photons and lead to the optical absorption in the visible range increases, which makes CrBr$_3$ with chalcogen doping to be possible candidates for optoelectronic applications.

\section{Methods}

Technical details of the calculations are as follows. All calculations of electronic structures and optical properties accurately were carried out using the Vienna ab initio simulation package (VASP) code~\cite{Kresse19961,Kresse19962} within the projector augmented-wave (PAW) method~\cite{Blochl1994,Kresse1999} and the exchange correlation functional of General gradient approximation (GGA) in the Perdew-Burke-Ernzerhof (PBE) implementation~\cite{Perdew1996}.
The thickness of the vacuum gap is at least 15 ${\rm \AA}$, which is large enough to avoid the interlayer interactions in the periodic structure.
By requiring convergence of results, the kinetic energy cutoff of $500$~eV and the Monkhorst-Pack $k$-mesh of 16$\times$16$\times$1 were used in all calculations about the electronic ground-state properties. The phonon spectra was calculated on a 4$\times$4$\times$1 $q$-grid using the density functional perturbation theory (DFPT)~\cite{Baroni2001} with VASP and Phonopy codes~\cite{Togo2008}. The biaxial strain was introduced by adjusting the lattice constant $a$ with the strain capacity $\varepsilon $= ($a$-$a_0$)/$a_0$$\times$100 \%, where $a_0$ is equilibrium lattice constant. The equilibrium lattice constant of the monolayer CrBr$_3$ was found to be $a_0$= 6.43 ${\rm \AA}$.

\section{Results}

Now we discuss the possible lattice structures of Cr$_2$Br$_{6-x}$S$_x$ with $x$ = 1$\sim$3. First, the hexagonal honeycomb primitive cell of CrBr$_3$ has eight atoms with each Cr surrounded by six Br atoms [Fig.~\ref{fig:phonon}(a)]. Due to the equivalency of six Br atoms in the primitive cell, Cr$_2$Br$_5$S has only one structure, as shown in Fig.~\ref{fig:phonon}(b). For the case of two S atoms, there are three different structures, marked by Cr$_2$Br$_4$S$_2$-A [Fig.~\ref{fig:phonon}(c)], Cr$_2$Br$_4$S$_2$-B [Fig.~\ref{fig:phonon}(d)] and Cr$_2$Br$_4$S$_2$-C (see Appendix A) for simplicity, corresponding to the adjacency, alternation and para-position in the top-view of Cr$_2$Br$_{6-x}$S$_x$, respectively. And there are two structures for the case of three S atoms. One is shown in Fig.~\ref{fig:phonon}(e) with three S atoms locating at the top and bottom edges of Cr atoms, Cr$_2$Br$_3$S$_3$-A. The other one have three ipsilateral S atoms, Cr$_2$Br$_3$S$_3$-B (see Appendix A), similar to the Janus monolayer structure of transition metal dichalcogenide~\cite{Lu2017}. The doping of S atom lead to the slight decrease of equilibrium lattice constant, as shown in Tab.~\ref{tab:bandgap}. Furthermore, the calculations of phonon spectra are applied to ensure the stability. The absent of imaginary frequency in the phonon spectrum of Cr$_2$Br$_6$ illustrates the dynamic stability of prototype structure, consistent with the previous works~\cite{Zhang2015}. After the doping of S atom, it is found that not all cases have stability. Cr$_2$Br$_5$S, Cr$_2$Br$_4$S$_2$-A, Cr$_2$Br$_4$S$_2$-B and Cr$_2$Br$_3$S$_3$-A are stable as shown in Fig.~\ref{fig:phonon}, contrary to the other cases (see Appendix A).

\begin{figure}[htp!]
\centerline{\includegraphics[width=0.5\textwidth]{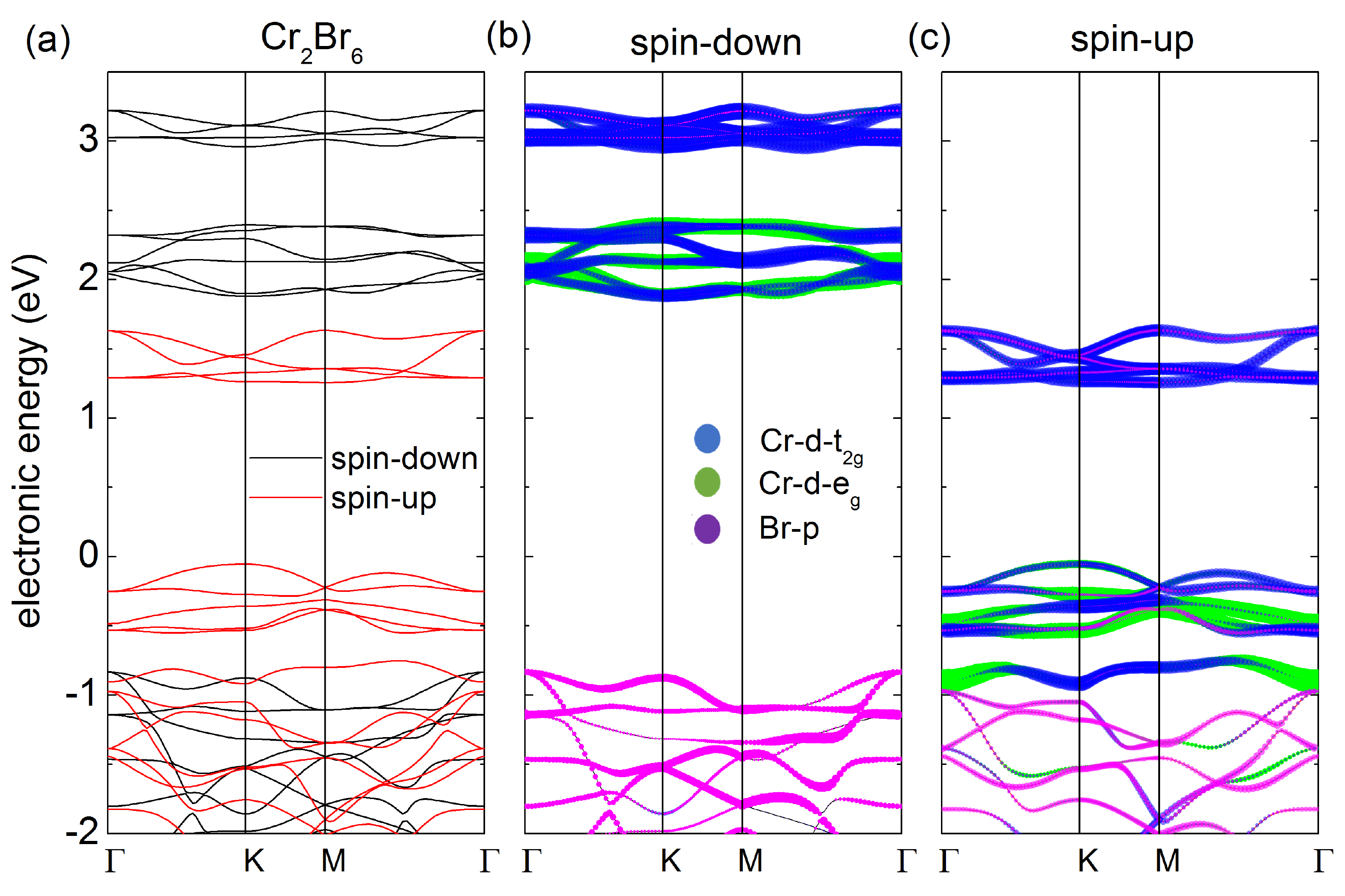}}
\caption{{\bf Band structures of Cr$_2$Br$_6$.} (a) Spin-polarized band structure of Cr$_2$Br$_6$. The red (black) lines represent the band structure in the spin-up (spin-down) direction. (b) Projected band structures in the spin-down direction and (c) spin-up direction of Cr-t$_{2g}$ (blue circle), Cr-e$_g$ orbits (green circle) and Br-p orbit (purple circle).
\label{fig:cbband}}
\end{figure}

\begin{figure}[tp!]
\centerline{\includegraphics[width=0.5\textwidth]{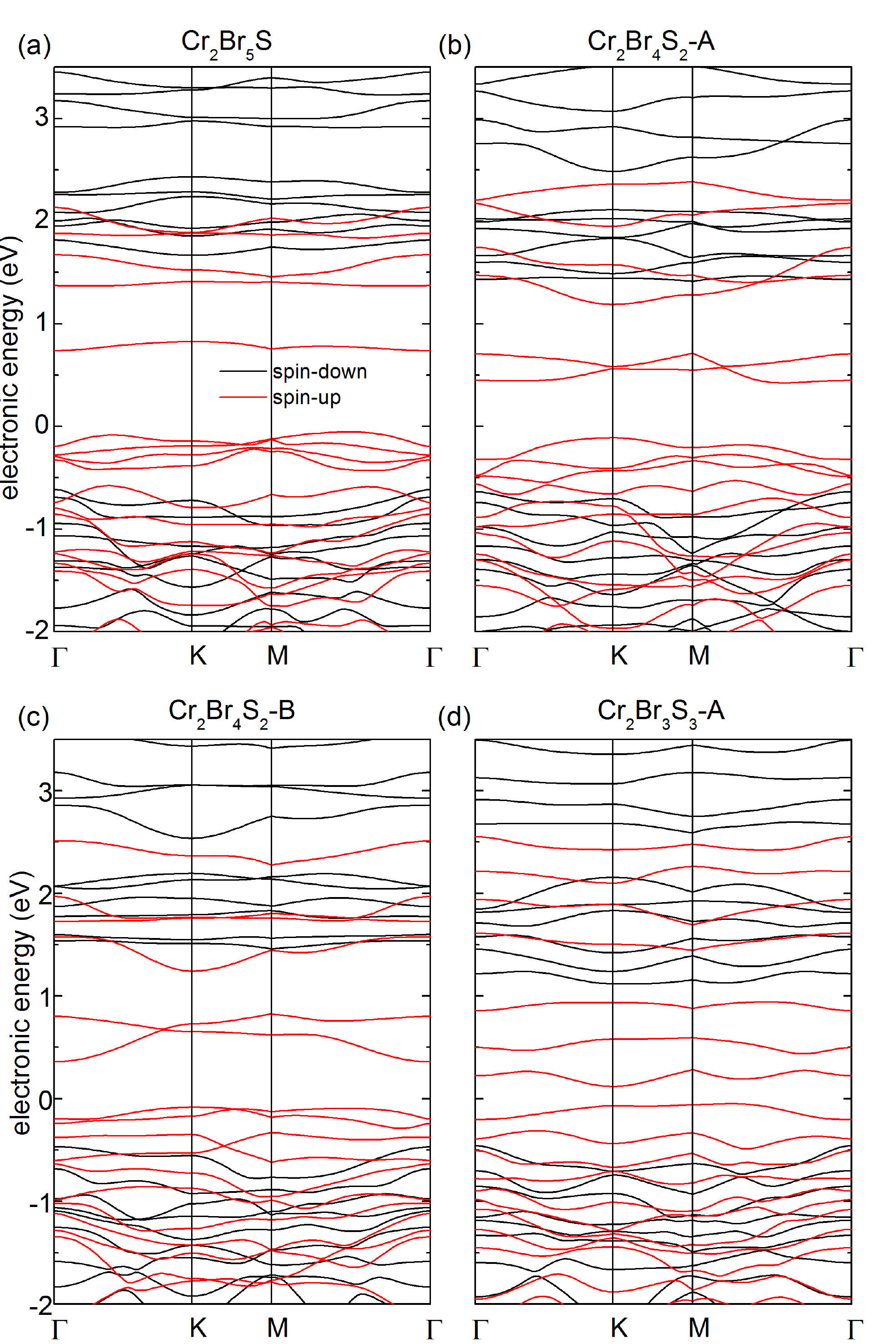}}
\caption{{\bf Band structures of Cr$_2$Br$_{6-x}$S$_x$.} (a) Spin-polarized band structure of Cr$_2$Br$_5$S, (b) Cr$_2$Br$_4$S$_2$-A, (c) Cr$_2$Br$_4$S$_2$-B and (d) Cr$_2$Br$_3$S$_3$-A. It's pretty obvious that the band gap decreases with the increase of S atoms.
\label{fig:cbsband}}
\end{figure}

The results of electronic structure show that monolayer Cr$_2$Br$_6$ is ferromagnetic semiconductor with the bandgap of 1.32 eV [Fig.~\ref{fig:cbband}(a)], in coordination with the previous work~\cite{Zhang2015}. The conduction band minimum (CBM) of band structure in the spin-down direction mainly consist of the d-t$_{2g}$ orbit of Cr atom as well as p orbit of Br atom contributing the valence band maximum (VBM) [Fig.~\ref{fig:cbband}(b)]. For the case of spin-up direction [Fig.~\ref{fig:cbband}(c)], CBM are also d-t$_{2g}$ orbit, but VBM is composed of d-t$_{2g}$ and d-e$_g$ orbits of Cr atom, clearly different from the spin-down direction. And the band structure of spin-up direction provides the global bandgap, much lower than that in spin-down direction [Tab.~\ref{tab:bandgap}]. Due to the ferromagnetic semiconductor, integral magnetic moment is also obtained with the value of 6 $\mu_{\rm B}$ per primitive cell [Tab.~\ref{tab:bandgap}], which are from 3 $\mu_{\rm B}$ of each Cr atom.

\begin{table}[htp!]
\caption{The characteristic parameter of electronic structure.}
\begin{tabular*}{8.5cm}{@{\extracolsep{\fill}}ccccccccc}
\hline\hline
        & $a_0$ & bandgap$_{\rm up}$ & bandgap$_{\rm down}$ & magnetic moment \\
        &  ${\rm \AA}$     & (eV) &  (eV) &  ($\mu_{\rm B}$/\ primitive cell) \\
\hline   Cr$_2$Br$_6$        &  6.43 & 1.32   &  2.72  &  6.0  & \\
         Cr$_2$Br$_5$S       &  6.37 & 0.79   &  2.29  &  5.0  & \\
         Cr$_2$Br$_4$S$_2$-A &  6.31 & 0.54   &  2.05  &  4.0  & \\
         Cr$_2$Br$_4$S$_2$-B &  6.32 & 0.44   &  1.93  &  4.0  & \\
         Cr$_2$Br$_3$S$_3$-A &  6.28 & 0.17   &  1.60  &  3.0  & \\
         Cr$_2$Br$_5$Se      &  6.36 & 0.79   &  2.27  &  5.0  & \\
\hline \hline
\end{tabular*}
\label{tab:bandgap}
\end{table}

\begin{figure}[htp!]
\centerline{\includegraphics[width=0.5\textwidth]{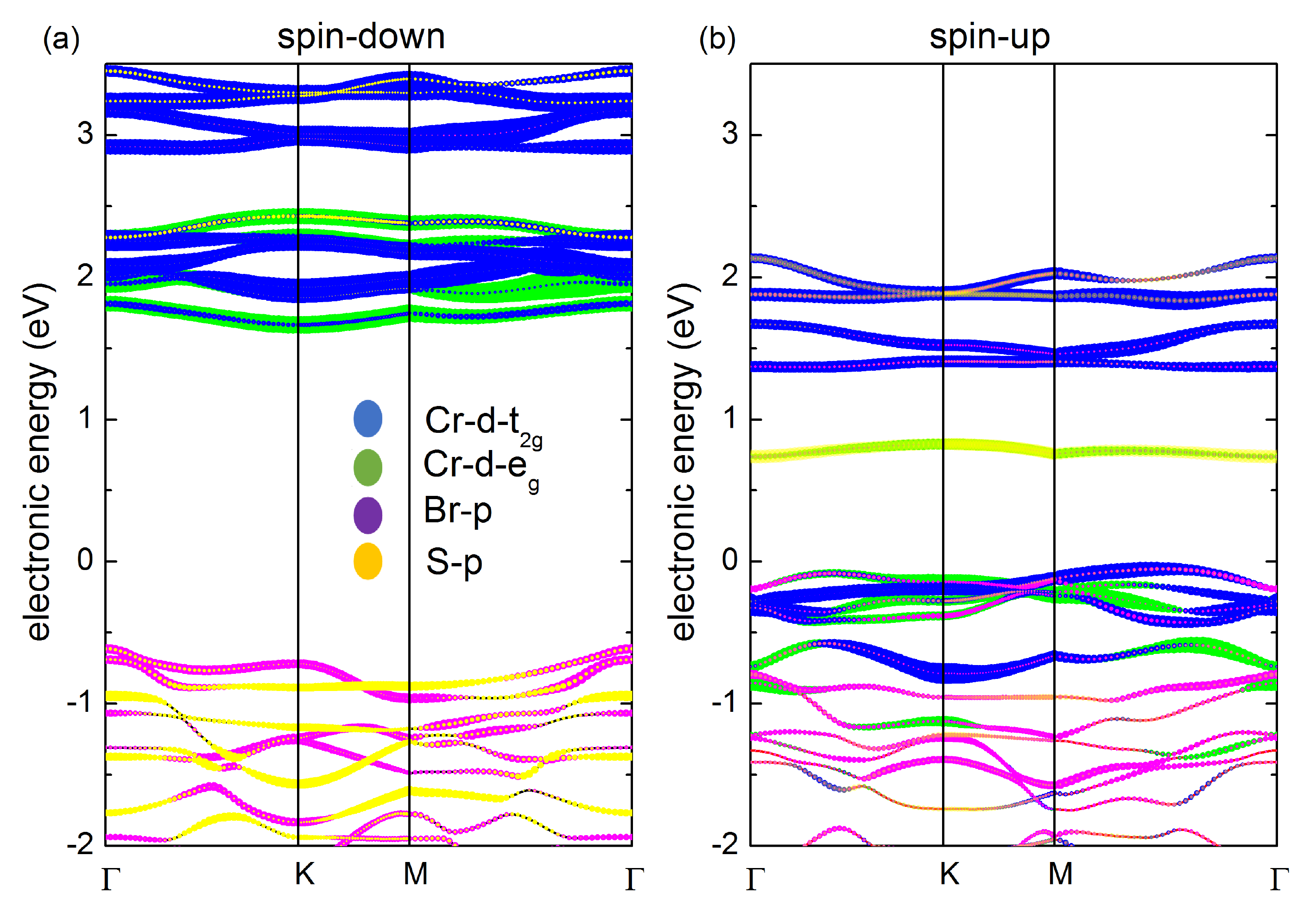}}
\caption{{\bf Band structures of Cr$_2$Br$_5$S.} (a) Projected band structures in the spin-down direction and (c) spin-up direction of Cr-t$_{2g}$ (blue circle), Cr-e$_g$ orbits (green circle), Br-p orbit (purple circle) and S-p orbit (yellow circle).
\label{fig:s1band}}
\end{figure}

We then study the doping effect of S atom on the electronic structure of Cr$_2$Br$_6$. All the four stable structures of Cr$_2$Br$_{6-x}$S$_x$ have the ferromagnetic ground state as same as Cr$_2$Br$_6$ and the ferromagnetic band structure are shown in Fig.~\ref{fig:cbsband}. It is obviously found that the global bandgap, spin-up direction, decreases rapidly with the increase in numbers of substitutional S atoms and reduces to a fairly low value of 0.17 eV in Cr$_2$Br$_3$S$_3$-A. The reduce of bandgap also occurs in the case of spin-down direction, as summarized in Tab.~\ref{tab:bandgap}. The projected band structures of Cr$_2$Br$_5$S show that S atom primarily influences the band structure in the spin-up direction [Fig.~\ref{fig:s1band}]. A new energy band presents in the energy gap between of Cr-d-t$_{2g}$ orbits and is contributed by the p orbit of S atom and d-e$_g$ orbit of Cr atom. The new bands in the energy range of 0 $\sim$ 1 eV for other cases are all the results from the same reason. And the effect of S atoms is also reflected by the magnetic moment, which decreases linearly with increase of S atoms. For example, there are still 3 $\mu_{\rm B}$ per Cr$_2$Br$_{3}$S$_3$-A primitive cell [Tab.~\ref{tab:bandgap}]. The main reason is that the decrease of valence electrons after S substitutions impact the occupation numbers on the magnetic Cr-d orbit. The results suggest that monolayer Cr$_2$Br$_{6-x}$S$_x$ are also robust intrinsic ferromagnetic semiconductor with different magnetic moments.

\begin{figure}[htp!]
\centerline{\includegraphics[width=0.5\textwidth]{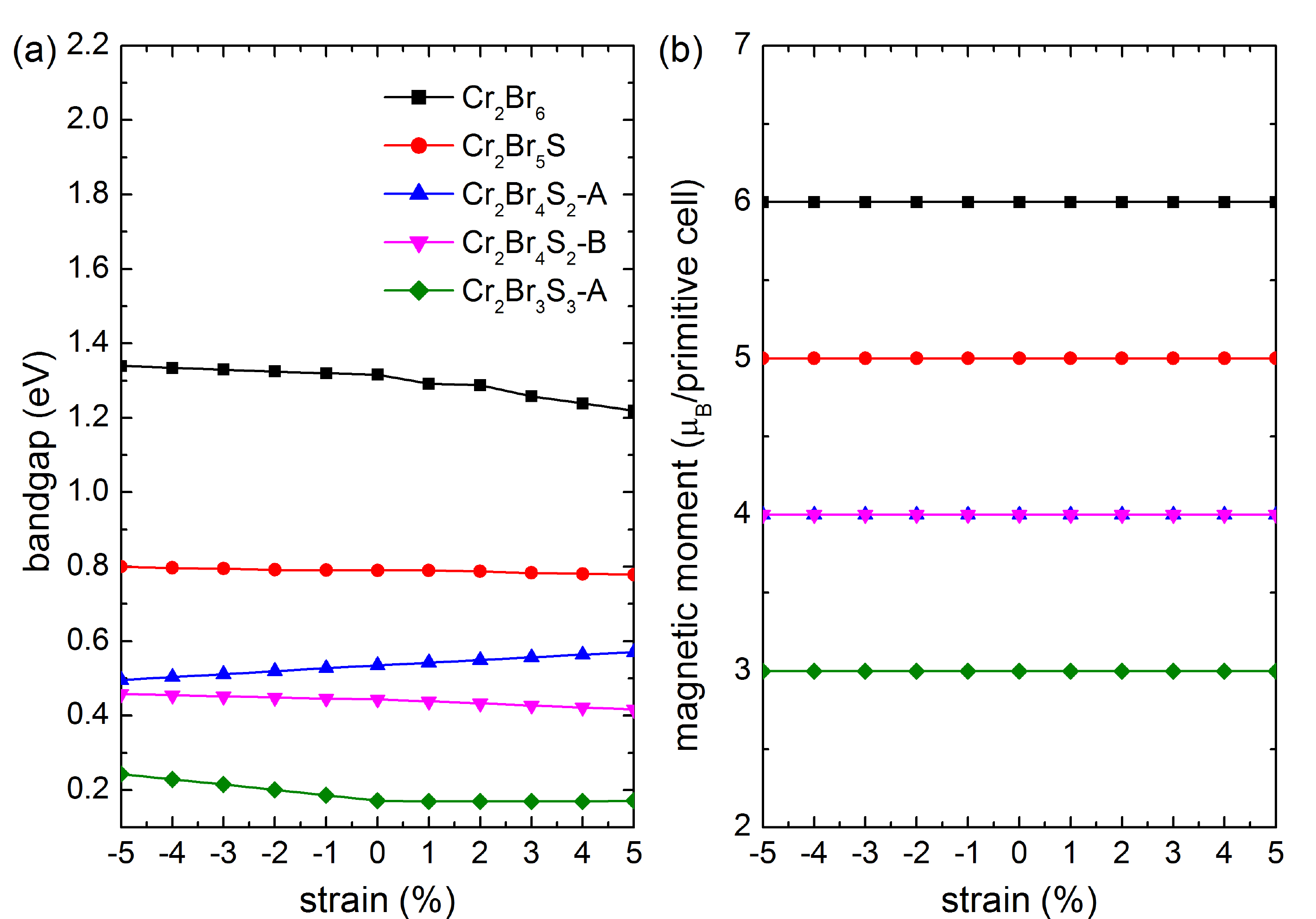}}
\caption{{\bf Strain effects.} (a) The strain dependence of bandgap and (b) magnetic moment of Cr$_2$Br$_{6-x}$S$_x$.
\label{fig:strain}}
\end{figure}
\begin{figure}[htp!]
\centerline{\includegraphics[width=0.5\textwidth]{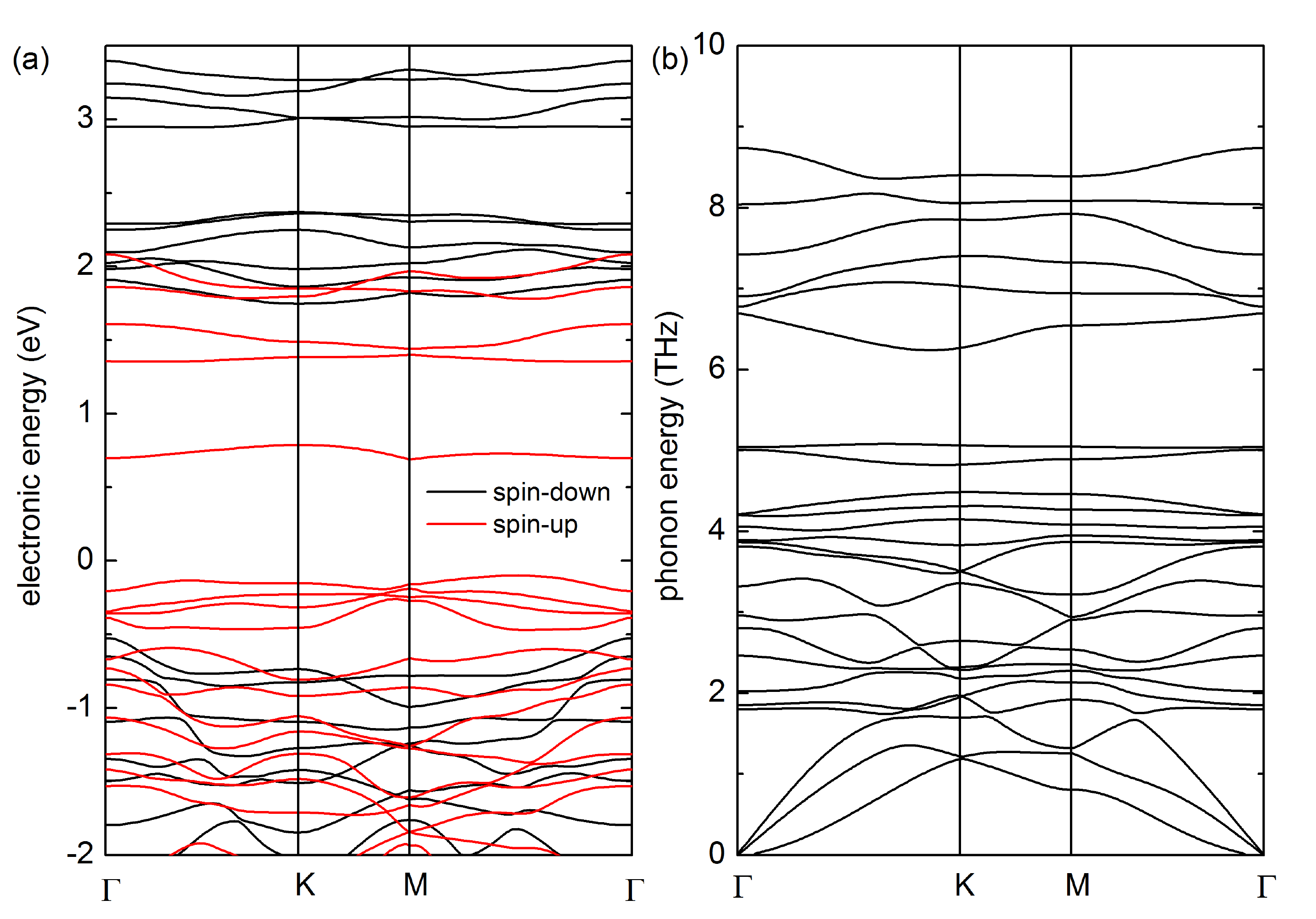}}
\caption{{\bf Band structure and phonon spectrum of Cr$_2$Br$_5$Se.}
\label{fig:se}}
\end{figure}

Because the possible lattice mismatching with the substrates in the different experimental preparations lead to the present of strain inevitably, the strain is also considered to study the stability of ferromagnetic semiconductor and the modulation of strain. The variation of bandgap under biaxial strains is displayed in Fig.~\ref{fig:strain}(a). The bandgap of Cr$_2$Br$_{6-x}$S$_x$ is robust to the biaxial strain obviously, and the bandgap of Cr$_2$Br$_5$S has the minimum change. The result of magnetic moment also illustrates that Cr$_2$Br$_{6-x}$S$_x$  maintains the character of ferromagnetic semiconductor in a wide range of strain. Beyond S atom, we also consider the effect of heavier chalcogen (Se and Te) in the same structures as above. But it is found that only Cr$_2$Br$_5$Se has structural stability, as shown in Fig.~\ref{fig:se}, while imaginary frequency exists in other situations, as summarized in Tab.~\ref{tab:frequency}. The electronic structures show that Cr$_2$Br$_5$Se is also ferromagnetic semiconductor and has a bandgap close to Cr$_2$Br$_5$S. Similarly, the new band in the range of 0 $\sim$ 1 eV also contributes from the Se-p and Cr-d-e$_g$ orbits.

\begin{table}[b!]
\caption{Imaginary frequency in unit of THz of the unstable structures.}
\begin{tabular*}{8cm}{@{\extracolsep{\fill}}ccccccccc}
\hline\hline
       &                      &  S        & Se        & Te \\
\hline &  Cr$_2$Br$_5$X       &  stable   &  stable   &  -0.09   \\
       &  Cr$_2$Br$_4$X$_2$-A &  stable   &  -0.05    &  -0.08   \\
       &  Cr$_2$Br$_4$X$_2$-B &  stable   &  -0.04    &  -0.12   \\
       &  Cr$_2$Br$_4$X$_2$-C & -0.32     &  -0.03    &  -0.12   \\
       &  Cr$_2$Br$_3$X$_3$-A &  stable   &  -0.07    &  -0.09   \\
       &  Cr$_2$Br$_3$X$_3$-B & -0.06     &  -0.08    &  -1.60   \\
\hline \hline
\end{tabular*}
\label{tab:frequency}
\end{table}

\begin{figure}[htp!]
\centerline{\includegraphics[width=0.5\textwidth]{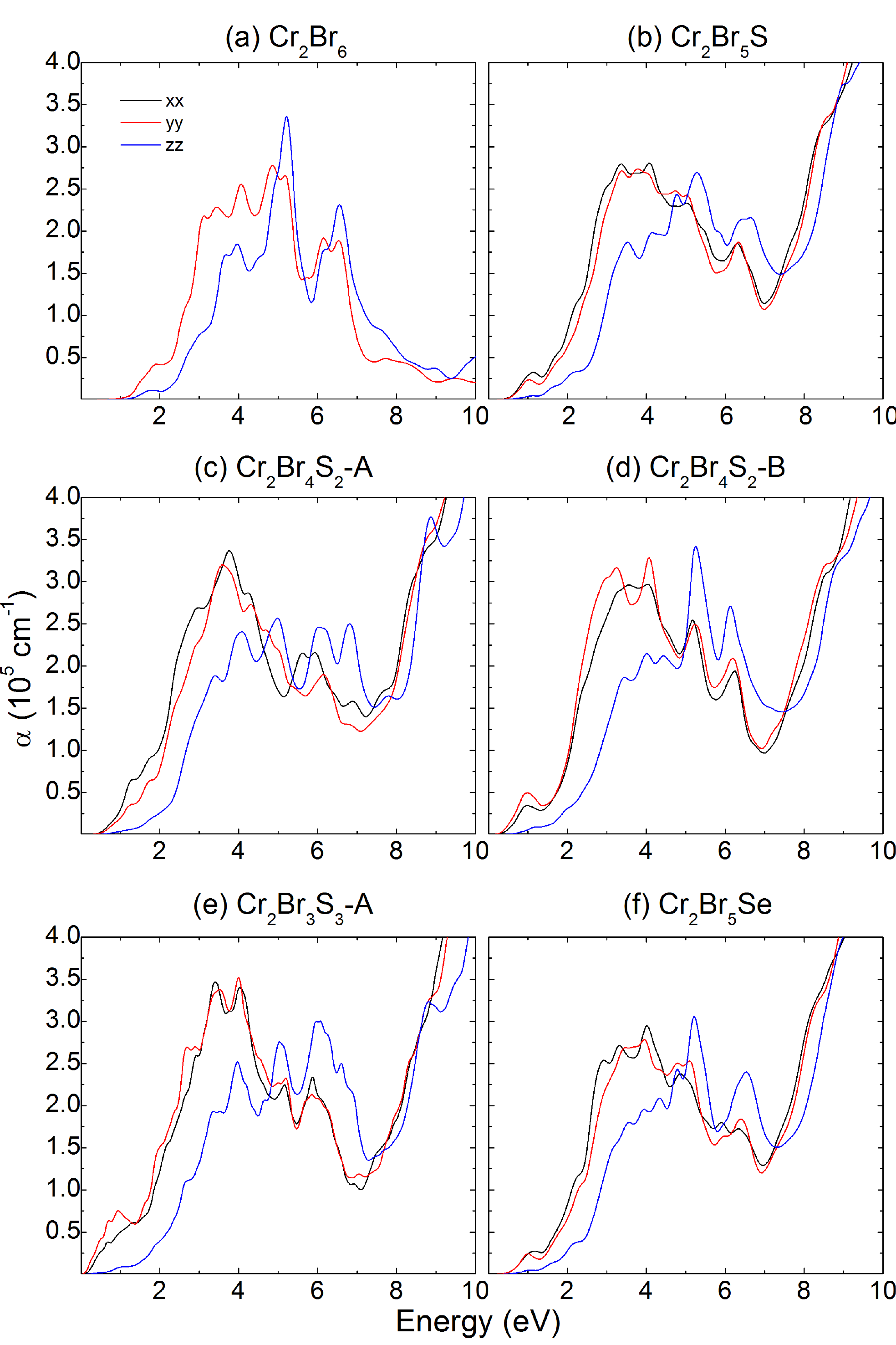}}
\caption{{\bf Optical absorption spectra.} (a) Cr$_2$Br$_6$, (b) Cr$_2$Br$_5$S$_1$, (c) Cr$_2$Br$_4$S$_2$-A, (d) Cr$_2$Br$_4$S$_2$-B, (e) Cr$_2$Br$_3$S$_3$-A and (f) Cr$_2$Br$_5$Se$_1$.
\label{fig:optic}}
\end{figure}

In addition, we have also calculated the doping effect on the optical properties of Cr$_2$Br$_6$ with chalcogen. As shown in Fig.~\ref{fig:optic}(a), the in-plane isotropic optical absorption spectra of xy plane distinct from that along the z axis because of the hexagonal symmetry of Cr$_2$Br$_6$. The onset peak around 1.3 eV in Cr$_2$Br$_6$ can be traced back to the excitation from valence band to lowest conduction band in the spin-up subband [Fig.~\ref{fig:cbband}(c)]. And the most obvious result after the doping is the break of in-plane isotropy, as shown in Fig.~\ref{fig:optic}(b-f), due to the position deviation of S or Se atom relative to Br atom in prototypical Cr$_2$Br$_6$ [Fig.~\ref{fig:phonon}]. Assisted by the new energy band in the original bandgap of spin-up subband, electronic transition can absorb the low-energy photons, so that the optical absorption of all doping cases in the visible range increases, especially Cr$_2$Br$_4$S$_2$-A and Cr$_2$Br$_3$S$_3$-A. This illustrates that Cr$_2$Br$_6$ with S and Se doping are possible candidates for optoelectronic applications.

\section{Discussion}

We have investigated the doping effect of chalcogen (S, Se and Te) on the CrBr$_3$ using first-principles calculations. We now take Cr$_2$Br$_{6-x}$S$_x$ as the example to summarize our results. Four stable configurations are predicted to be ferromagnetic semiconductors: Cr$_2$Br$_5$S, Cr$_2$Br$_4$S$_2$-A, Cr$_2$Br$_4$S$_2$-B and Cr$_2$Br$_3$S$_3$-A. After the doping of S atom, the new bands appearing in the energy range of 0 $\sim$ 1 eV are made up of S-p and Cr-d-e$_g$ orbits and lead to the obvious reduce of bandgap in the spin-up direction. Since the change of valence electron, the magnetic moment also decreases with the increase of S atoms, such as Cr$_2$Br$_3$S$_3$-A has the minimum magnetic moment of 3 $\mu_{\rm B}$ per primitive cell. And the character of ferromagnetic semiconductor is always hold in a wide range of strain. Furthermore, the doping of chalcogen can also lead to the increase of optical absorption in the visible range and make it to be possible candidates for optoelectronic applications.

As an important material in FM semiconductor, CrI$_3$ is considered certainly, but we do not find any stable structures, as shown in Appendix B.
Except for the substitution, we also calculate the case of atomic adds because of the vacancy at the lattice vertex [Fig.~\ref{fig:phonon}]. The large imaginary frequency also explain the structural instability of Cr$_2$Br$_{6}$S and Cr$_2$Br$_{6}$Se (see Appendix C).
For the unstable structures, especially the structures with very small imaginary frequency, such as the Janus Cr$_2$Br$_3$S$_3$-B and Cr$_2$I$_5$S, maybe prepared in the experiments with the help of substrate, which can enhance 2D material stability. Importantly, we have shown that monolayer CrBr$_3$ with chalcogen doping is also robust intrinsic ferromagnetic semiconductor and supply an effectual way to control the magnetism and optic properties of monolayer CrBr$_3$, which is the current experiments desperately needed.

\begin{acknowledgments}
This work was supported by the NSFC (Grants No.11747054 and No.11874113), the Specialized Research Fund for the Doctoral Program of Higher Education of China (Grant No.2018M631760), the Project of Heibei Educational Department, China (No. ZD2018015 and QN2018012), and the Advanced Postdoctoral Programs of Hebei Province (No.B2017003004).
\end{acknowledgments}

\setcounter{figure}{0}
\appendix
\section{Unstable structures of Cr$_2$Br$_{6-x}$S$_x$}

\begin{figure}[ht!]
	\renewcommand\thefigure{A\arabic{figure}}
	\centerline{\includegraphics[width=0.5\textwidth]{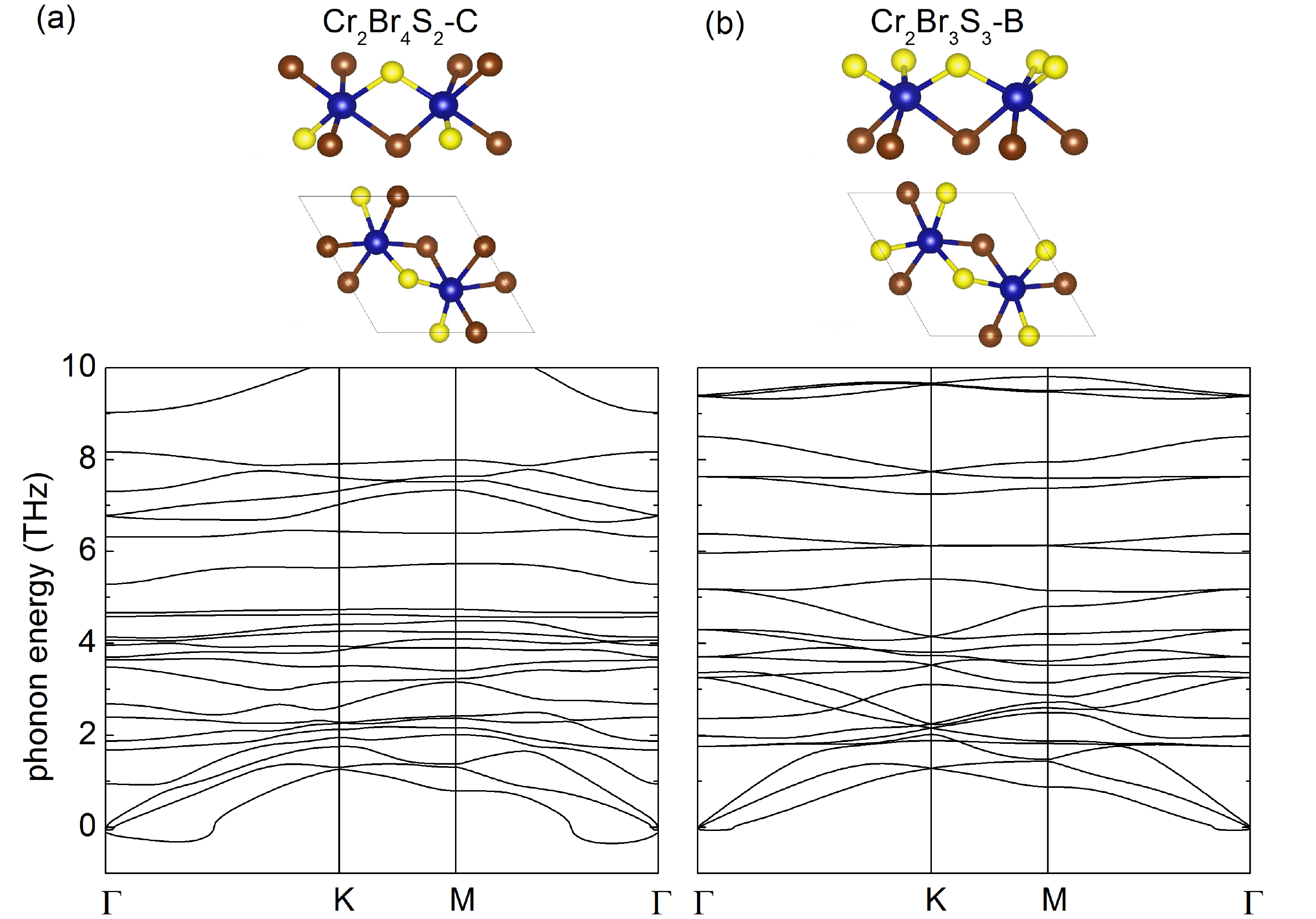}}
	\caption{{\bf Crystal structures and phonon spectra of Cr$_2$Br$_4$S$_2$-C and Cr$_2$Br$_3$S$_3$-B.}
		\label{fig:aphonon}}
\end{figure}

It is found that two of the crystal structures Cr$_2$Br$_{6-x}$S$_x$ considered in the present work are unstable, ensured by the imaginary frequency around $\Gamma$ point, as shown in Fig.~\ref{fig:aphonon}. And the value of imaginary frequency in Cr$_2$Br$_3$S$_3$-B is smaller than that of Cr$_2$Br$_4$S$_2$-C as well as the distribution of imaginary frequency.

\section{Cr$_2$I$_{6-x}$S$_x$}

\begin{figure}[ht!]
	\renewcommand\thefigure{A\arabic{figure}}
	\centerline{\includegraphics[width=0.5\textwidth]{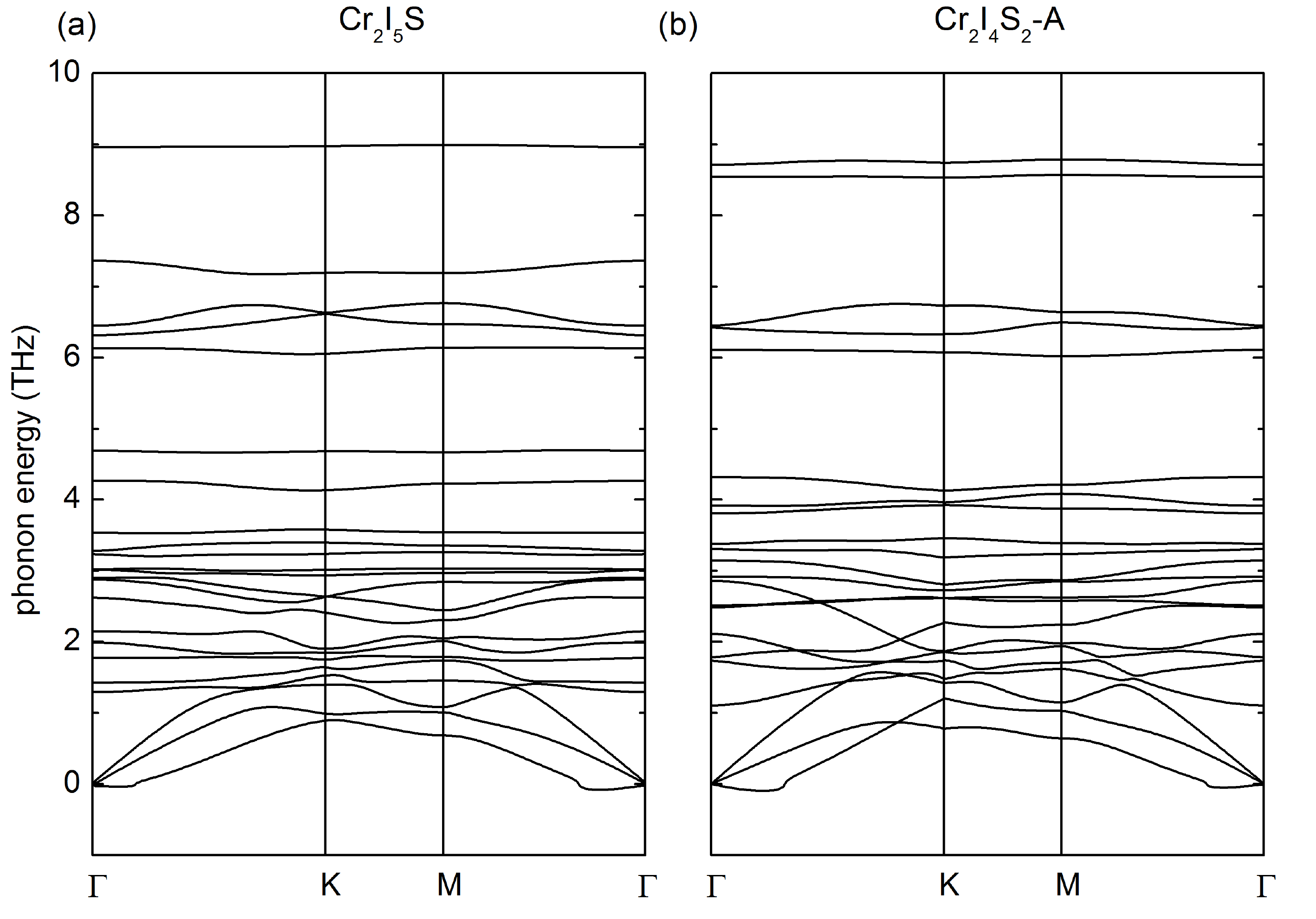}}
	\caption{{\bf Phonon spectra of Cr$_2$I$_5$S and Cr$_2$I$_4$S$_2$-A.}
		\label{fig:cri3s}}
\end{figure}

The doping effect of chalcogen is also calculated in the CrI$_3$ system. But we haven't obtained any stable structure and figure.~\ref{fig:cri3s} only plots the phonon spectra of Cr$_2$I$_5$S and Cr$_2$I$_4$S$_2$-A with small imaginary frequency around $\Gamma$ point.

\section{Cr$_2$Br$_{6}$S and Cr$_2$Br$_{6}$Se}

\begin{figure}[ht!]
	\renewcommand\thefigure{A\arabic{figure}}
	\centerline{\includegraphics[width=0.5\textwidth]{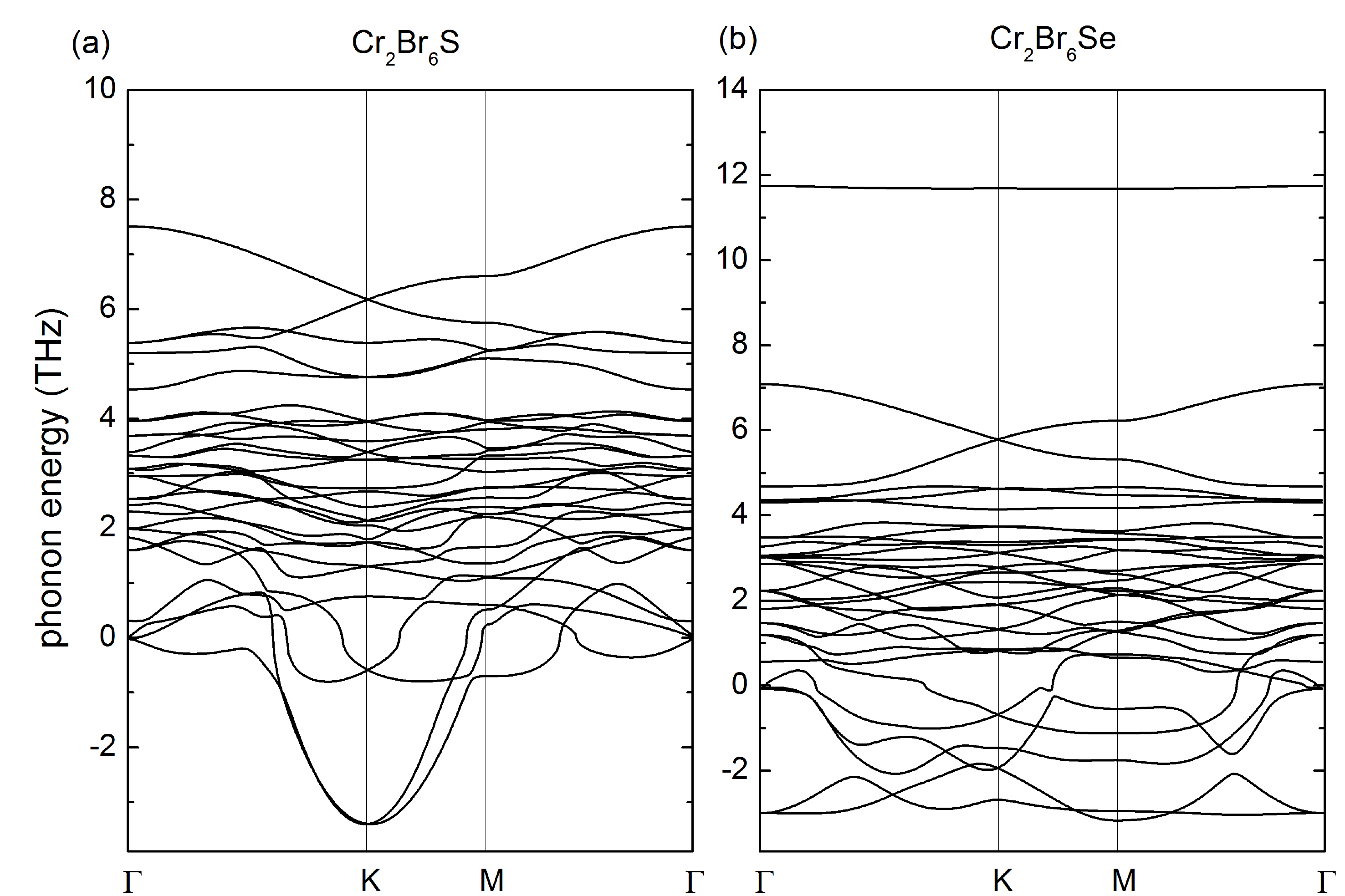}}
	\caption{{\bf Phonon spectra of Cr$_2$Br$_6$S and Cr$_2$Br$_6$Se.}
		\label{fig:add}}
\end{figure}

The addition of chalcogen is also calculated in the CrBr$_3$ system. But there are large imaginary frequency in the phonon spectra, as shown in Fig.~\ref{fig:add}.

\end{document}